\documentclass[preprint,aps,nofootinbib]{revtex4}

\usepackage{dcolumn}
\usepackage{bm}
\usepackage{epsfig}
\usepackage{graphicx,subfigure}

\usepackage{amsmath}
\usepackage{amssymb}
\usepackage{color}
\usepackage[usenames,dvipsnames]{xcolor}
\usepackage{hyperref}

\def\fun#1#2{\lower3.6pt\vbox{\baselineskip0pt\lineskip.9pt
  \ialign{$\mathsurround=0pt#1\hfil##\hfil$\crcr#2\crcr\sim\crcr}}}

\input epsf

\newcommand{\be}{\begin{equation}}
\newcommand{\ee}{\end{equation}}
\newcommand{\bea}{\begin{eqnarray}}
\newcommand{\eea}{\end{eqnarray}}

\begin{document}

\begin{flushright}
\vspace*{-1.5cm}
\vspace{-0.2cm}ANL-HEP-PR-12-85\\
\vspace{-0.2cm}EFI-12-31\\
\vspace{-0.2cm}FERMILAB-PUB-12-614-T\\
\vspace{-0.2cm}MCTP-12-29\\
\vspace{-0.2cm}NSF-KITP-12-209
\end{flushright}

\vspace*{0.1cm}

\title{Vacuum Stability and Higgs Diphoton Decays in the  MSSM}

\vspace*{0.2cm}

\author{
\vspace{0.2cm} 
\mbox{\bf Marcela Carena$^{\,a,b,c}$, Stefania Gori$^{\,b,d}$,}\\
\mbox{\bf Ian Low$^{\,d,e,f}$, Nausheen R. Shah$\,^g$, and Carlos E.~M.~Wagner$^{\,b,c,d}$}
 }
\affiliation{
\vspace*{.4cm}
$^a$  \mbox{Fermi National Accelerator Laboratory, P.O. Box 500, Batavia, IL 60510}\\
$^b$  \mbox{Enrico Fermi Institute, University of Chicago, Chicago, IL 60637}\\
$^c$  \mbox{Kavli Institute for Cosmological Physics, University of Chicago, Chicago, IL 60637}\\
$^d$ \mbox{High Energy Physics Division, Argonne National Laboratory, Argonne, IL 60439}\\
$^e$ \mbox{Department of Physics and Astronomy, Northwestern University, Evanston, IL 60208} \\
$^f$\mbox{Kavli Institute for Theoretical Physics, University of California, Santa Barbara, CA 93106}\\
$^g$\mbox{Michigan Center for Theoretical Physics, Department of Physics,} \\
\mbox{University of Michigan, Ann Arbor, MI 48109}\\
}

\begin{abstract}
\vspace*{0.3cm}
Current Higgs data at the Large Hadron Collider is compatible with a SM signal at the 2$\sigma$ level, but the central value of the signal strength in the diphoton channel is enhanced with respect to the SM expectation. If the enhancement resides in the diphoton partial decay width, the data could be accommodated in the Minimally Supersymmetric Standard Model (MSSM) with highly mixed light staus. We revisit the issue of vacuum instability induced by large mixing in the  stau sector, including effects of a radiatively-corrected tau Yukawa coupling. Further, we emphasize the importance of taking into account the $\tan\beta$ dependence in the stability bound. While the metastability of the Universe constrains the possible enhancement in the Higgs to diphoton decay width in the light stau scenario, an increase of the order of 50\%  can  be achieved in the region of  large $\tan\beta$. Larger enhancements may be obtained, but would require  values  of $\tan\beta$  associated with non-perturbative values of the tau Yukawa coupling at scales below the GUT scale, thereby implying the presence of new physics beyond the MSSM. 
\end{abstract}

\maketitle

\section{Introduction}
The discovery of a new boson at the CERN Large Hadron Collider (LHC) is an extraordinary achievement for high energy physics \cite{:2012gk,:2012gu}. Preliminary studies indicate that the properties of the new boson are generally consistent with that of a Standard Model (SM) Higgs boson \cite{Low:2012rj}. It is of course crucial to eventually measure the spin, CP, and electroweak quantum numbers of this new particle. In current data the significance of the discovery is mainly driven by the observation of an excess of events in the diphoton and four-lepton channels, that  are compatible at the 2 $\sigma$ level with a SM-like Higgs boson with a mass at around 125 GeV. 

Additionally, there are  indications that the new particle decays into $WW$ with SM-like rates. Searches are also being performed in the $b\bar{b}$ and $\tau\tau$ channels. The decay rates in these channels also show consistency with SM expectations at the 1~$\sigma$ level~\cite{Aaltonen:2012qt},\cite{LHCbottomtau}. Although the diphoton signal is compatible with a SM Higgs signal at the 2 $\sigma$ level, the central value of the signal strength is observed to be approximately 2 and 1.5 times the SM prediction by the ATLAS and CMS collaborations,  respectively~\cite{gagamu}. Taken at face value, the data seems to suggest that the enhancement in the  diphoton rate arises from an enhanced partial decay width of the Higgs to diphotons, while the  production cross-sections seem to indicate that the couplings of the new resonance to vector bosons and top quarks are similar to those predicted for the SM Higgs. At present, the measurements are statistically limited, and more data is necessary to reach conclusive results.

The deviation in the diphoton rate from the SM prediction, if confirmed, would be a clear indication of physics beyond the SM. In the MSSM, the  best  motivated candidates that can significantly  modify the Higgs to diphoton decay width are  sleptons, in particular  staus~\cite{Carena:2011aa, Carena:2012gp}. A significant enhancement requires that one of the staus is light, of the order of 100 GeV, and further that the mixing between the left- and right-handed staus is large. 

More generally, an increase in the Higgs to diphoton partial width without impacting the production rate requires new color neutral, electrically charged particles with masses of the order of 100 GeV and significant couplings to the Higgs boson~\cite{Carena:2011aa}--\cite{NPinhiggsloops}. Such new particles  inevitably modify the Higgs potential through quantum corrections. If the new charged particles are fermions, they may drive the Higgs quartic coupling negative via renormalization group evolution and thus destabilize the Higgs vacuum at scales of the order of a few TeV~\cite{Joglekar:2012vc}. For new charged scalars, an enhanced diphoton width requires a large cubic coupling between the Higgs and a pair of the new scalars, which in turn could induce a new charge-breaking vacuum \cite{Carena:2012xa}. It is therefore important to study the extent to which the diphoton partial width could be enhanced via light charged particles without inducing  vacuum instability   at energies well below the GUT scale.

In the MSSM, constraints from charge-breaking minima induced by large stau mixing were studied in Ref.~\cite{Rattazzi:1996fb}, and later refined in Ref.~\cite{Hisano:2010re}. More recently,  Refs.~\cite{Sato:2012bf,Kitahara:2012pb}  considered vacuum stability issues in the heavily mixed, light stau scenario with enhanced Higgs to diphoton partial width based on the results in Ref.~\cite{Hisano:2010re}. In this work we revisit the vacuum stability issue and improve upon the work in Ref.~\cite{Hisano:2010re}. In particular, we emphasize the importance of  retaining the dependence on the ratio of the Higgs vacuum expectation values,  $\tan\beta$,  for fixed values of  the stau mixing mass parameter, $\mu \tan\beta$, when deriving vacuum stability bounds. We also include important $\tan\beta$-enhanced effects from radiatively corrected tau Yukawa couplings. Our study shows that a diphoton partial width enhancement of  ${\cal O}$(50\%) above the SM expectation can be compatible with the metastability of the electroweak-breaking vacuum for sufficiently large  $\tan \beta$ in the MSSM. An enhancement beyond ${\cal O}$(50\%) requires such large values of $\tan\beta$  that  the $\tau$ Yukawa coupling  becomes non-perturbative below the GUT scale, which in turn would imply new physics beyond the MSSM at scales below the corresponding Landau pole.
  
This work is organized as follows. In Sec.~\ref{sect:II} we analyze the dependence of the vacuum stability conditions on the parameters of the model, such as  Yukawa couplings, $A_\tau$, and $m_A$,  and compare with previous results. We further make connection with the enhancement of the Higgs to diphoton decay width with these same parameters. In Sec.~\ref{sec:Deltatau}, we investigate the radiative corrections to the tau and bottom Yukawa couplings and their impact on the vacuum stability conditions. Sec.~\ref{sec:HiggsCoup} discusses the effects of $A_\tau$ and $m_A$ on the total width of the Higgs decay,  which enters into the diphoton decay branching fraction. Numerical results, which follow from these considerations,  are presented  in Sec.~\ref{sec:results}. We then discuss some of the possible constraints on the large $\tan \beta$ region in Sec.~\ref{sec:tb}. In Sec.~\ref{sec:Conc} we present  our conclusions.

\section{staus,  Higgs diphoton width  and Vacuum Stability}
\label{sect:II}
 
To study the metastability of the electroweak vacuum in the MSSM, in the presence of light  staus with large mixing, it is instructive to first write down the scalar potential for the neutral component of the up-type Higgs, $h_u$, the left-handed stau, $\tilde{\tau}_L$, and the right-handed stau, $\tilde{\tau}_R$. 
 Neglecting the down-type Higgs is a very good approximation since  for large values of  $\tan\beta$ - as 
required  to achieve large mixing in the stau sector- and sizable $m_A$,  the $h_d$ vacuum expectation value~(VEV) remains small. However, for completeness, when presenting most  of our numerical results in Section \ref{sec:results}, we will use the full potential including both the up- and the down-type Higgs bosons. 

Following Ref.~\cite{Rattazzi:1996fb}, and normalizing all fields as complex scalar fields, the scalar potential can be written as
\bea
\label{eq:potential}
V&=& \left| \mu\, h_u - y_\tau \tilde{\tau}_L \tilde{\tau}_R^* \right|^2 + \frac{g_2^2}{8}\left(|\tilde{\tau}_L|^2+|h_u|^2\right)^2+\frac{g_1^2}{8}\left(|\tilde{\tau}_L|^2-2|\tilde{\tau}_R|^2-|h_u|^2 \right)^2  \nonumber \\
&& + m_{H_u}^2 |h_u|^2 + {m}_{L_3}^2 |\tilde{\tau}_L|^2 + {m}_{E_3}^2 |\tilde{\tau}_R|^2 + \frac{g_1^2+g_2^2}{8}\, \delta_H\, |h_u|^4 \ ,
\eea
where $\mu$ is the Higgsino supersymmetric mass parameter and $y_\tau$ is the tau Yukawa coupling appearing in the MSSM superpotential. $g_2$ and $g_1$ are the gauge couplings for $SU(2)_L$ and $U(1)_Y$, respectively. In addition, $m_{H_u}^2$, ${m}_{L_3}^2$, and ${m}_{E_3}^2$ are the soft-breaking masses for the up-type Higgs, the left-handed third generation sleptons, and the right-handed third generation sleptons. The last term, proportional to $\delta_H$ in Eq.~(\ref{eq:potential}), represents the leading contribution to the full one-loop effective potential, arising from the top and stop loops. This contribution depends on the average stop mass, $m_{\tilde{t}}$, and on the stop mixing parameter, $X_t=A_t-\mu\cot \beta$,  and is  of the order of unity for a 125 GeV Higgs boson~\cite{Hisano:2010re},  
\be
\delta_H^{(t)} = \frac{3}{2 \pi^2} \frac{y_t^4}{g_1^2+g_2^2}\left[ \log\left( \frac{m^2_{\tilde{t}}}{m_t^2} \right) + \frac{X_t^2}{m^2_{\tilde{t}}} - \frac{X_t^4}{12 m_{\tilde{t}}^4} \right] \  \sim\ 1 \ ,
\ee
where $y_t \approx \sqrt{2} m_t/v$ with  $v\approx 246$ GeV, and $m_t$ is the weak scale running top-quark mass.

The source of vacuum instability in Eq.~(\ref{eq:potential}) is clear: the term  coupling the Higgs to the two staus, whose coefficient is proportional to $\mu\, y_\tau$, has a negative sign which tends to destabilize the vacuum for positive values of the fields. As first studied in Ref.~\cite{Rattazzi:1996fb}, when this  cubic coupling becomes too large, a charge breaking vacuum deeper than the electroweak breaking vacuum may exist. Moreover, after the Higgs acquires a VEV, $h_{u(d)}\to (v_{u(d)}+ h_{u(d))})/ \sqrt{2}$, this cubic coupling, $\mu\, y_\tau/\sqrt{2}$, also contributes to the off-diagonal entry in the stau mass-squared matrix. Including $\tan\beta$-suppressed terms that were not included in the potential in Eq. (\ref{eq:potential}), the stau mass matrix is given by 
\be
\label{eq:staumass}
{\cal M}_{\tilde{\tau}}^2 = \left( \begin{array}{cc}
          m_{L_3}^2+m_\tau^2+ D_L &  \frac{y_\tau v}{\sqrt{2}} \, (A_\tau \cos\beta -\mu \sin\beta) \\
         \frac{ y_\tau v }{\sqrt{2}} \, (A_\tau\cos\beta -\mu \sin\beta) & m_{E_3}^2+m_\tau^2+ D_R
          \end{array} \right) \ ,
\ee   
where $A_\tau$ is the soft-breaking $A$-term for staus, and $D_{L,R}$ are the $D$-term contributions to the slepton masses. Therefore the coefficient that triggers the vacuum instability is also crucial in determining the mixing in the stau sector and hence the possible enhancement in the Higgs to diphoton width.

A charge breaking minimum is not necessarily a problem if the ordinary electroweak breaking vacuum is metastable with a lifetime longer than the age of the Universe. The lifetime of a metastable vacuum is usually computed using semiclassical techniques. The probability of decaying into the true vacuum per unit spacetime volume is given by \cite{Coleman:1977py}
\be\label{eq:tunneling}
\frac{\Gamma}{V} = A \, e^{- S_E} \ ,
\ee
where $A \sim (100 \ {\rm GeV})^4$ is a dimensionful parameter, expected to be roughly the fourth power of the electroweak scale. The $S_E$ is the Euclidean action evaluated on the ``bounce'' solution that interpolates between the metastable vacuum and the other side of the barrier. The volume is given by $V=R^4$, with $R$ being the characteristic size of the bounce. Asking for the lifetime of the metastable vacuum to be longer than the present age of the Universe  is equivalent to requiring  $\Gamma /V $ 
to be smaller than the fourth power of the Hubble constant, $H_0 \sim 1.5\times 10^{-42}$ GeV. This then implies that the vacuum is metastable if  \cite{Claudson:1983et,Kusenko:1996jn}
\be
S_E\  \agt\  400\ .
\ee

In Ref.~\cite{Hisano:2010re} a numerical study based on Eq.~(\ref{eq:potential}) in the configuration space of the three fields $h_u, \tilde{\tau}_L$ and $\tilde{\tau}_R$, found that the metastability condition is mainly sensitive to $\mu\tan\beta$, $m_{{L}_3}$, and $m_{{E}_3}$, while the dependence on $\tan\beta$  for fixed $\mu\tan\beta$ and on $\delta_H$ is small. The resulting vacuum metastability condition in Ref. \cite{Hisano:2010re} was summarized as follows\footnote{We recently learned that Eq.~(\ref{eq:hbound}) is being revised \cite{private}.\label{ft:1}}
\be
\label{eq:hbound}
|\mu\tan\beta| < 76.9 \sqrt{m_{{L}_3} m_{E_3}} + 38.7 (m_{{L}_3}+m_{{E}_3}) - 1.04 \times 10^4 \ {\rm GeV}\ .
\ee
The small dependence of the metastability condition on $\delta_H$, as claimed in Ref.\cite{Hisano:2010re},  is understandable since it only affects the Higgs quartic term. In particular, the bound on $\mu\tan\beta$ changes by only $\sim 10\%$ when  $\delta_H$ is varied from 0 to 1. In our study, we will not investigate this dependence further and, unless otherwise stated, fix $\delta_H$ at 0.9 since the mass of the Higgs has been measured to be approximately $125$ GeV. 

The dependence on $\tan\beta$ can be understood starting from the tree-level relation for the tau lepton mass and the tau Yukawa coupling,
\be
\label{eq:treeytau}
y_\tau \approx \sqrt{2}\,  \frac{m_\tau}{v \cos\beta}\ .
\ee
When $\tan\beta\gg 1$, $\sin\beta\approx 1-1/(2\tan^2\beta)$, which very quickly approaches unity, hence we can approximate $\sin\beta\approx 1$. Then Eq.~(\ref{eq:treeytau}) implies that for $\tan\beta \gg 1$,  the dependence of $y_\tau$ on $\tan\beta$ is, to a very good approximation
\be
\label{eq:ytautanbeta}
y_\tau \approx \sqrt{2} \, \frac{m_\tau}{v} \frac{\tan\beta}{\sin\beta} \approx  \frac{\tan\beta}{100} \ .
\ee 
As mentioned earlier, the coefficient of the destabilizing cubic term in the scalar potential, $h_u\tilde{\tau}_L\tilde{\tau}_R$, is proportional to $\mu y_\tau$, and therefore to $\mu\tan\beta$. However, there is also a term which depends only on  $\tan^2\beta$  which  arises from  the stabilizing quartic term, $|\tilde{\tau}_L\tilde{\tau}_R|^2$, whose coefficient is $y_\tau^2$. Therefore, for fixed  $\mu\tan\beta$, or equivalently fixed  $\mu y_\tau$, there is a residual dependence on $\tan \beta$ coming from this stabilizing term.

Using the constraint in  Eq.~(\ref{eq:hbound}) from Ref.~\cite{Hisano:2010re}, that directly relates the vacuum metastability condition to the stau mixing in the MSSM, the author of  Ref.~\cite{Kitahara:2012pb} derived a maximum allowed enhancement of the Higgs to diphoton width of about 25\% of the SM value, for a light stau mass heavier than about 100~GeV. Our work, however, shows that the bound on $\mu \tan\beta$ is about 15\% larger than the one displayed in Eq.~(\ref{eq:hbound}). Moreover,  we will show that the residual dependence on $\tan \beta$, for a fixed $\mu \tan \beta$, can have a significant impact on the  vacuum metastability requirement. This is to be expected since if $\mu\tan\beta$ (or equivalently $\mu y_\tau$) is held constant, then going to larger values of $\tan\beta$ increases the coefficient of the stabilizing quartic term, $|\tilde{\tau}_L\tilde{\tau}_R|^2$, and alleviates the vacuum stability constraint. This implies that larger values of $\tan\beta$ will lead to a further relaxation of the bound on $\mu\tan\beta$, allowing for larger enhancement of the Higgs to diphoton rate from light stau loops. As we will show below, within the MSSM the magnitude of the possible enhancement is thus constrained by how large the value of $\tan\beta$ (or equivalently of $y_\tau$  at low energies) can be, without implying that the tau Yukawa coupling develops a   Landau Pole   at energies below the GUT scale. 

We shall make comparisons with previous works, such as Ref.~\cite{Hisano:2010re}, using the improved tree-level approximation in Eq.~(\ref{eq:potential}). However, our final results are obtained using the full one-loop effective potential, including also terms involving either $h_d$ or suppressed by $\tan \beta$ which were neglected in Eq.~(\ref{eq:potential}). Including such terms, in the large $\tan \beta$ limit, we obtain  additional contributions to the scalar potential:
\begin{eqnarray} 
 \label{eq:deltaV}
 \Delta V & \simeq  & m_A^2 |h_d|^2 - \frac{m_A^2}{\tan\beta}( h_d h_u + {\rm h.c.}) + \frac{m_A^2}{\tan^2\beta} |h_u|^2 + (y_\tau A_\tau h_d \tilde\tau_L \tilde\tau_R^* + {\rm h.c.} ) 
 \nonumber\\
  & + & |y_\tau|^2 |h_d|^2 \left(|\tilde\tau_L|^2 + |\tilde\tau_R|^2\right) + D\mbox{-terms}.
 \end{eqnarray}
Comparing this with Eq.~(\ref{eq:potential}), we see that $m_A$ and $A_\tau$ could also impact  vacuum stability.  First we note that  for positive values of $A_\tau$ and $\mu$, in the charge breaking minima,  the field $h_d$ tends to acquire values which are opposite in sign to the $h_u$ values. This means that positive,  non-negligible values of $A_\tau$ contribute constructively to the destabilizing trilinear term in the scalar potential, thereby tightening the bound on $\mu\tan\beta$  from the metastability condition. However, the $m_A^2$ terms still give a positive contribution to the scalar potential. Therefore, the $A_\tau$ effects are suppressed for large values of $m_A$, which then tend to reduce the values of  $h_d$ associated with the effective potential minima. Hence we see that additional charge breaking minima may  be induced for small values of $m_A$ and large values of $A_\tau$. This in turn implies that  non-zero $A_\tau$ with $\mu A_\tau >$ 0, can further suppress the possible enhancement of the diphoton width coming from light staus, depending on the value of $m_A$. 
 
We are interested in regions of parameter space where  $A_\tau$ is smaller than or of the order of 1 TeV and $\mu \tan \beta \sim \mathcal{O}(30)$ TeV. In this region, $A_\tau$ does not directly play  an important role in the value of the stau mass, as can be seen from Eq.~(\ref{eq:staumass}). However, for a given set of parameters, a positive value of $\mu A_\tau$  lowers the minimum value of $\mu \tan \beta$ allowed by the metastability condition, suppressing the mixing effect in the stau sector. This then increases the lightest stau mass compatible with vacuum stability and reduces the enhancement in the diphoton partial decay width. On the other hand, positive values of  $\mu A_\tau$ reduce the total width of the lightest CP-even Higgs via mixing in the CP-even Higgs sector \cite{Carena:2011aa,Carena:2012gp}. It turns out that the reduction in the enhancement of the diphoton partial width is largely compensated by the decrease in the total decay width, leaving a diphoton branching fraction that is not severely affected by $A_\tau$. The effect on the total width of the Higgs due to mixing effects in the Higgs sector will be discussed in detail in Sec.~\ref{sec:HiggsCoup}.

\subsection{Radiative Corrections to the Tau and Bottom Yukawas}
\label{sec:Deltatau}

As stated in Ref.~\cite{Hisano:2010re}, the reason the stabilizing effect from $\tan\beta$ is  naturally suppressed is because the tree-level tau Yukawa coupling, $y_\tau$, is proportional to $\tan\beta$ with a small proportionality constant  $\sim1/100$, Eq.~(\ref{eq:ytautanbeta}), and only very large values of $\tan \beta$ would overcome that suppression.  The proportionality constant  in Eq.~(\ref{eq:treeytau}) is modified at one-loop level to be \cite{Carena:2002es}
\be
\label{eq:looptau}
y_\tau = \sqrt{2}\,  \frac{m_\tau}{v \cos\beta (1+\Delta_\tau)}\approx  \frac{\tan\beta}{100 (1+\Delta_\tau)}\ ,
\ee
where $\Delta_\tau$ arises dominantly from a stau-neutralino  and a sneutrino-chargino loop, and may become of $\mathcal O(0.1)$ for sufficiently large values of $\tan\beta$. An approximate expression for $\Delta_\tau$ is given in the Appendix in Eq.~(\ref{eq:deltatau}). The full analytic expression for $\Delta_\tau$ can be found in Ref.~\cite{Carena:2002es} and has been implemented in a new version of the code {\tt CPsuperH} \cite{cpsuperh}, which we use in our study. 

The bottom Yukawa coupling receives similar modifications,  
\be
\label{eq:loopb}
y_b = \sqrt{2}\,  \frac{m_b}{v \cos\beta (1+\Delta_b)} \ ,
\ee
where $\Delta_b$ is dominated by contributions from sbottom-gluino  and  stop-chargino loops. An approximate expression for $\Delta_b$ is also give in the Appendix. From Eq.~(\ref{eq:deltab}) we see that, for stop masses of the order of 1~TeV and  sizable trilinear terms $A_t$ that are needed to accommodate a 125 GeV Higgs, the stop-chargino loop contribution becomes sizable. So, while the sbottom-gluino loop contribution could be suppressed by a large mass splitting between the sbottom and the gluino, the stop-chargino loop contribution in $\Delta_b$ is always sizable in the scenario we consider~\footnote{Alternatively, one could have stop and sbottom masses of the order of  a few tens of TeV  to obtain a 125 GeV Higgs without a significant $A_t$. In that case, both the gluino-sbottom as well as the stop-chargino loop would be suppressed and $\Delta_b$ would be small.}.

The $\Delta_\tau$ corrections are smaller in magnitude compared to the $\Delta_b$ corrections because they are suppressed by electroweak gauge couplings. Additionally, $\Delta_\tau$  tends to be dominated by loops which include electroweak gauginos and acquires a sign opposite to that of $\mu M_2$. On the other hand,  for squark and gluino masses that are of the same order, the sbottom-gluino contribution to $\Delta_b$ becomes the  dominant one and $\Delta_b$  acquires the same sign as $\mu M_3$. Furthermore, the stop-chargino contribution is proportional to $\mu A_t$ and adds to the effect of the sbottom-gluino loop if $\mu A_t$ has the same sign as $\mu M_3$. In the following we shall  consider values of $\mu M_i > 0$ for all three gaugino masses, $i =1, 2, 3$, and $\mu A_t > 0$, which then lead to positive values of $\Delta_b$ and negative values of $\Delta_\tau$. Note that this choice of signs improves the agreement between the theoretical prediction of the anomalous magnetic moment of the muon
and its measured value~\cite{Bennett:2006fi},\cite{g-2} and also helps in weakening the bounds on the value of $\tan\beta$ coming from the requirement of keeping the bottom Yukawa coupling perturbative until high energy scales. Moreover, as we shall discuss in more detail below, positive values of $\mu A_t$ are helpful in avoiding the constraints coming from the $B_s \to \mu\mu$ rare decay measurement~\cite{LHCb}.

From Eq.~(\ref{eq:looptau}) it is clear that one can define an effective $\tan\beta$ in the tau sector~\footnote{Notice that one could define a different effective $\tan\beta$ in the bottom sector.} by
\be
\label{eq:tanbetaeff}
\tan\beta_{\tau\, {\rm eff}} \equiv \frac{\tan\beta}{1+\Delta_\tau} \ ,
\ee
which simplifies the relation between $y_\tau$ and $m_\tau$. Looking at the stau mass matrix, given in Eq.~(\ref{eq:staumass}), we see that if we further define an effective $A_\tau$ by
\be
\label{eq:ataueff}
A_{\tau\,{\rm eff}} \equiv \frac{A_\tau}{1+\Delta_\tau} \ ,
\ee
then  the stau mass-squared matrix can be re-written as
\be
\label{eq:1loopstaumass}
{\cal M}_{\tilde{\tau}}^2 = \left( \begin{array}{cc}
          m_{L_3}^2+m_\tau^2+ D_L & m_\tau  (A_{\tau\,{\rm eff}}  -\mu \tan\beta_{\tau\, {\rm eff}}) \\
          m_\tau  (A_{\tau\,{\rm eff}}  -\mu \tan\beta_{\tau\, {\rm eff}}) & m_{E_3}^2+m_\tau^2+ D_R
          \end{array} \right) \ .
\ee  

Since the relation between $m_\tau$ and $y_\tau$, as well as the stau mass-squared matrix, retain their tree-level form when using the effective $\tan\beta$ defined in Eq.~(\ref{eq:tanbetaeff}), we will find it convenient to express our results in terms of this effective quantity. 

\begin{figure}[t]
\includegraphics[scale=0.5, angle=0]{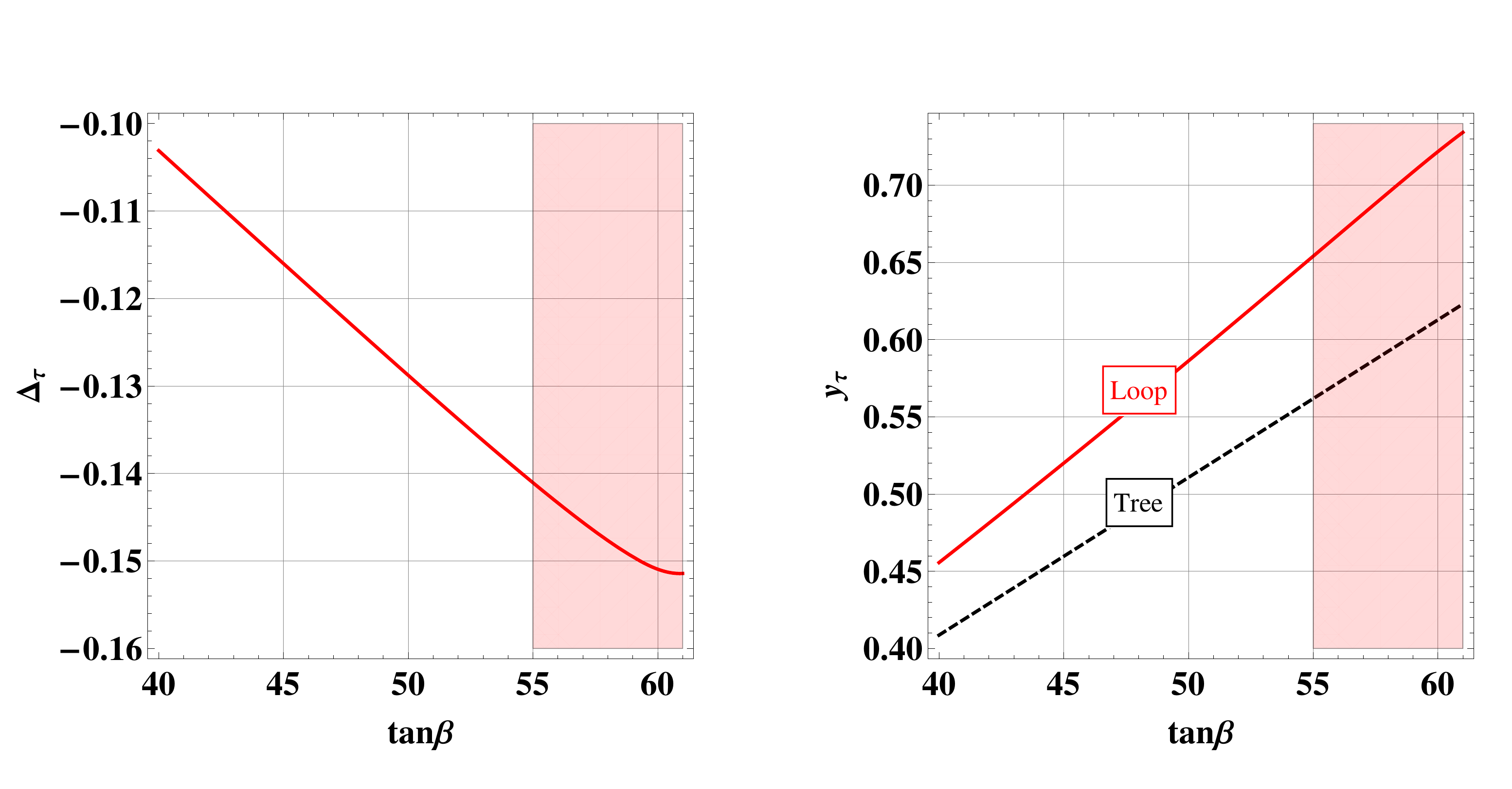} 
\caption{\em $\Delta_\tau$ versus $\tan\beta$ (left panel) and  $y_\tau$ versus $\tan\beta$ (right panel)  for $m_{L_3}=m_{E_3}=250$ GeV, $\mu=520$ GeV  and $A_\tau=m_A=1$ TeV. In the right panel, the red (solid) line uses the one-loop relation, including $\Delta_\tau$, while the black (dashed) line uses the tree-level relation. Shaded regions in both figures correspond to a stau mass below the LEP bound of 90 GeV.}
\label{fig:deltatb} 
\end{figure}

 In Fig.~\ref{fig:deltatb} we plot $\Delta_\tau$ and $y_\tau$ as functions of $\tan\beta$ for $m_{L_3}=m_{E_3}=250$ GeV, $\mu=520$ GeV, $M_2 = 400$~GeV and $A_\tau=1$ TeV. The shaded region in the figures corresponds to a stau mass that is below the LEP limit of about 90 GeV~\cite{staulimit}. From the left panel we see that $\Delta_\tau$ is negative and of the order of 10 -- 15\% for $\tan\beta = 40$--60.  The right panel of Fig.~\ref{fig:deltatb} shows the effect of $\Delta_\tau$ in increasing the value of $y_\tau$ for any given value of $\tan\beta$.  For example, for $\tan\beta=50$ the corresponding $y_\tau$ value,  without  the  inclusion of  the $\Delta_\tau$ effect, can be read-off using the line labeled  ``Tree"  to be $\approx 0.51$. Including $\Delta_\tau$ increases the associated value of $y_\tau$ to be the one read from the line labeled ``Loop", giving $\approx 0.585$. In addition, for this particular choice of $m_{L_3}, m_{E_3}, \mu$, $A_\tau$ and $M_2$, this figure also provides a translation between $\tan\beta$ and $\tan\beta_{\tau\, {\rm eff}}$, as defined in Eq.~(\ref{eq:tanbetaeff}). For instance, as stated above, the value of the radiatively corrected $y_\tau$ associated with $\tan \beta=50$ is $\sim 0.585$. The $\tan\beta_{\tau\, {\rm eff}}$ corresponding to this $y_\tau$ can then be read-off to be $\approx 57.5$ using the line labeled ``Tree"~(which is simply the relationship defined in Eq.~(\ref{eq:treeytau})). 

From the above analysis, it is clear that including the effect of $\Delta_\tau$ allows for a  larger tau Yukawa coupling, $y_\tau$, which increases the values of the stau mixing parameter, $\mu y_\tau $, allowed by metastability constraints. Such large stau mixing effects, in turn, allow for a larger enhancement of the rate of the Higgs decay into diphotons.

\subsection{Higgs Couplings to $\tau\tau$ and $b\bar{b}$}
\label{sec:HiggsCoup}

As mentioned briefly before, smaller values of $m_A$  and sizable values of $A_\tau$ induce an additional new physics effect in the diphoton event rate associated with a reduced Higgs total decay width. In the MSSM, the lightest CP-even Higgs is a linear combination of $h_u$ and $h_d$. The mixing angle, $\alpha$, is governed by the off-diagonal element of the CP-even Higgs mass matrix, 
 \begin{equation}
 \label{eq:higgsmass}
{\cal M}_H^2  =
\left[
\begin{tabular}{c c}
$m_A^2 \sin^2\beta + M_Z^2 \cos^2\beta$          &  $-(m_A^2 + M_Z^2) \sin\beta \cos\beta +{\rm Loop}_{12}$ \\
$ -(m_A^2 + M_Z^2) \sin\beta \cos\beta +{\rm Loop}_{12}$         & $m_A^2 \cos^2\beta + M_Z^2 \sin^2\beta +{\rm Loop}_{22}$,
\end{tabular}
\right]\;,
\end{equation}
where~\cite{Carena:2002es,Carena:2011aa} 
\begin{equation}
{\rm Loop}_{12}  \simeq \frac{m_t^4 }{16 \pi^2 v^2 \sin^2\beta} \frac{\mu A_t}{M_{\rm SUSY}^2}  \left[  \frac{A_t^2}{M_{\rm SUSY}^2} - 6 \right]
+ \frac{y_b^4 v^2}{16 \pi^2} \sin^2\beta \frac{\mu^3 A_b}{M_{\rm SUSY}^4}  + \frac{y_{\tau}^4 v^2}{48 \pi^2} \sin^2\beta \frac{\mu^3 A_{\tau}}{M_{\tilde{\tau}}^4}\,,
\label{loop12}
\end{equation}
and 
\be
\sin(2 \alpha) = \frac{2 (\mathcal{M}_H^2)_{12}}{\sqrt{Tr[\mathcal{M}_H^2]^2-4\det[\mathcal{M}_H^2]}}\,.
\label{eq:sin2alpha}
\ee

At tree level, $h_u$ couples only to the up-type fermions and $h_d$ to the down-type fermions (leptons and down-type quarks). Since the lightest Higgs is given by the combination
 \be
 h = - h_d\, \sin\alpha +  h_u\, \cos\alpha \ ,
 \ee
 its tree-level coupling to down-type fermions are then given by
 \be
 \label{eq:ghddtree}
 g_{hd{d}} = -\frac{\sin\alpha}{\cos\beta} \, \frac{m_d}{v} \,.
 \ee
 Furthermore,  Eq.~(\ref{eq:ghddtree}) gets corrected at one-loop~\cite{Carena:2002es} :
\bea
g_{hdd} &=&  -\frac{\sin\alpha}{\cos\beta} \, \frac{m_d}{v} \frac{1}{(1 + \Delta_d)} \left( 1 - \frac{\Delta_d}{\tan\alpha \tan\beta} \right) 
\label{ghdd}
\eea

In the decoupling regime where $m_A$ is large, we have $\sin\alpha\to -\cos\beta$ and $\cos\alpha\to \sin\beta$, so that the lightest CP-even Higgs couplings to fermions approach their SM values, but significant departures from these values may be obtained for smaller values of $m_A$.  In the absence of loop corrections to the Higgs mass matrix elements, Eq.~(\ref{loop12}),  the down-type  fermion couplings to the Higgs tend to be enhanced with respect to the SM values for moderate or small values of $m_A$~\cite{patrick}.  However, for a non-zero and positive $\mu A_\tau$,  the loop-corrections may lead  to a relevant suppression of the off-diagonal term in the Higgs mass matrix in Eq.~(\ref{eq:higgsmass}) and consequently a reduction of $|\sin\alpha|$. As a result, the bottom and  $\tau$  couplings of the lightest CP-even Higgs boson may be also suppressed. The outcome is that the Higgs total width may be reduced and the  branching fractions into gauge boson pairs, including diphotons, will be enhanced.

Note that for large values of $A_\tau$, the suppression in the $h\tau\tau$ coupling is larger than the one in the $hbb$ coupling. This follows from the fact that the value of $\sin\alpha$ is small and negative and $\sin\alpha/\cos\beta \simeq \tan\alpha \tan\beta$. For $|\Delta_d| <1$, a reduction of the Higgs decay into down-type fermions may only be obtained for $|\sin\alpha/\cos\beta| < 1$. This then implies that the coefficient of $\Delta_d$ in the numerator of Eq.~(\ref{ghdd}) must be positive and larger than one. Therefore, a negative (positive) $\Delta_d$ would decrease (increase) the $g_{hdd}$ coupling compared to the case $\Delta_d=0$. Since in our scenario $\Delta_b$ is positive and $\Delta_\tau$ is negative, a sizable suppression of the $\tau$ coupling of the Higgs may be induced for large values of $\tan\beta$ and positive $A_\tau$, while the Higgs coupling to bottom quarks remains closer to the SM value.

\subsection{Results} 
 \label{sec:results}
 
In the following we will consider $\tan\beta_{\tau\,{\rm eff}}$ as an input parameter to study the interplay between the vacuum stability constraint and the possible enhancement of the Higgs to diphoton partial width.

\begin{figure}[t]
\includegraphics[scale=0.48, angle=0]{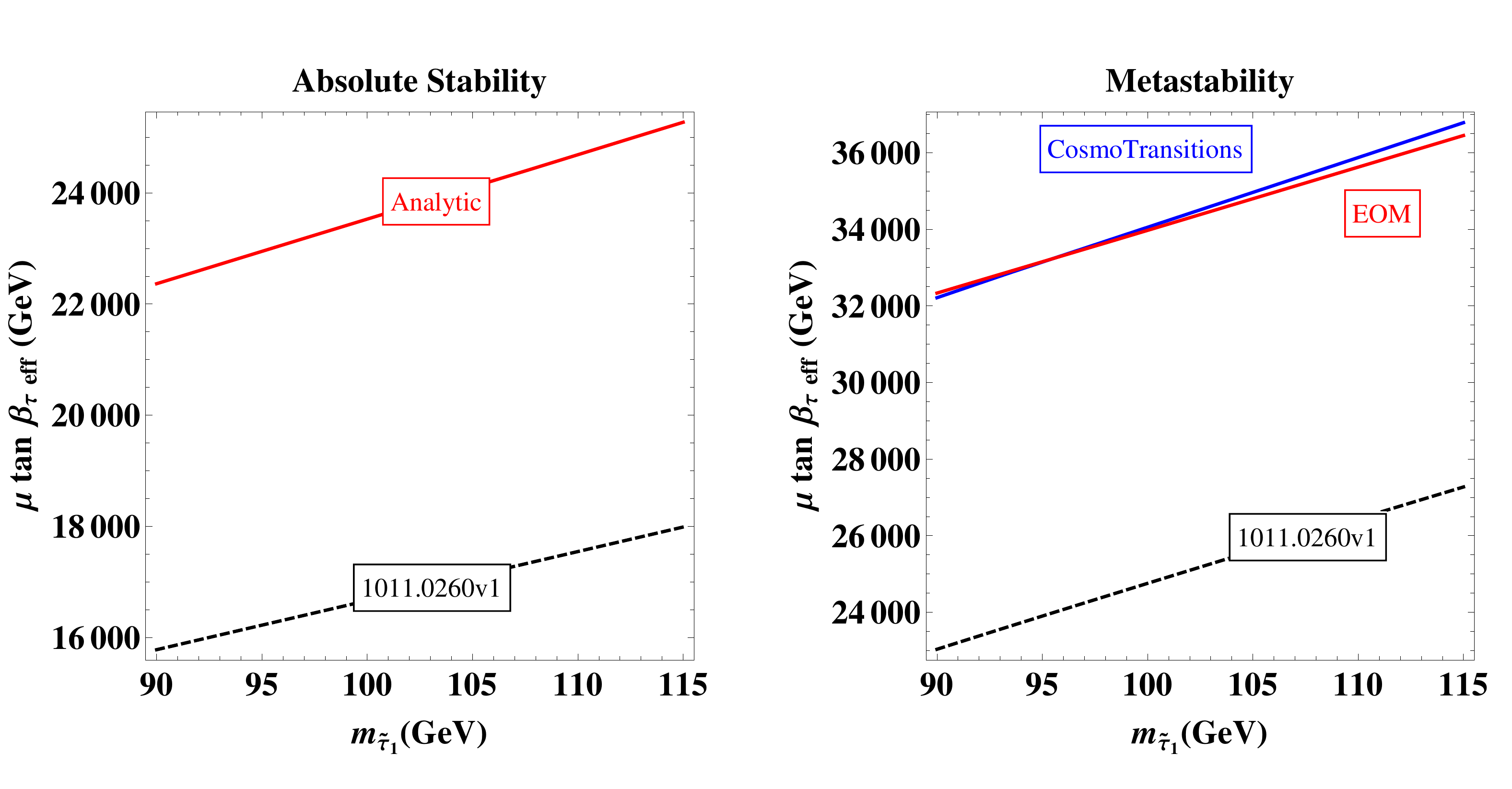} 
\caption{\label{fig:70}{\em Right panel: Metastability bound on $\mu\tan\beta_{\tau\,{\rm eff}}$ as a function of $m_{\tilde{\tau}_1}$, having fixed $A_{\tau}=0$, $m_A=2$ TeV and $\tan\beta_{\rm eff}=70$. The black dashed line is the bound obtained in Ref.~\cite{Hisano:2010re}, the blue solid line is the bound obtained by {\tt CosmoTransitions} and the red solid line is the bound obtained solving the one dimensional equations of motion (see text). Left panel: Analytic absolute stability bound on $\mu\tan\beta_{\tau\,{\rm eff}}$  as a function of $m_{\tilde{\tau}_1}$, for  the same set of supersymmetry parameters} as in the right panel, and the comparison with the bound from Ref.~\cite{Hisano:2010re}.}
\end{figure}

In Fig.~\ref{fig:70} we show the bound on $\mu\tan\beta_{\tau\,{\rm eff}}$ from vacuum stability constraints as a function of the lightest stau mass, for $A_{\tau}=0$, $m_A=2$~TeV, $\tan\beta_{\tau\,\rm eff}=70$ and $\delta_H=1$. In the left panel, we show the ``absolute stability" bound obtained by imposing that the electroweak minimum is the global minimum of the theory; in the right panel  we show the ``metastability" bound obtained by imposing that the electroweak minimum is only a local minimum but with a life time longer than the age of the Universe. The corresponding bounds from Ref.~\cite{Hisano:2010re} are shown by the black dashed lines. 
 
We compute the metastability bound using the following two methods: 
 
  \begin{itemize}
 
\item  The numerical package {\tt CosmoTransitions} \cite{cosmoT}, which uses a path deformation technique to compute the bounce solution for a multi-dimensional scalar potential. We refer the reader to  Ref.~\cite{Wainwright:2011kj} for  details. The result corresponding to the scalar potential in Eq.~(\ref{eq:potential}) is shown by the blue solid line.

\item  A numerical procedure which computes the bounce action by reducing the problem to a one dimensional one:  at large values of $\mu\tan\beta$, we define a canonically normalized  field, $\Phi$, which connects the charge breaking minimum and the electroweak breaking minimum. We compute the scalar potential of the new field, $V_1(\Phi)$, using the potential in Eq. (\ref{eq:potential}). In this one-dimensional setup, the bounce solution is then calculated using the conventional ``over-shoot/under-shoot" method \cite{Coleman:1978ae}. The red solid line shows this result.

 \end{itemize}
We emphasize that, in order to compare with Ref.~\cite{Hisano:2010re},  the results of the above two methods,  shown in the right panel of Fig.~\ref{fig:70}, are evaluated using the improved tree-level scalar potential in Eq.~(\ref{eq:potential}).  One sees that the metastability bounds obtained by the two methods we use  are in excellent agreement with each other, however, they are significantly less stringent than the bound from Ref.~\cite{Hisano:2010re}~\footnote{To confirm our results, we used a third method based on Ref.~\cite{Duncan:1992ai},  which approximates the one dimensional potential described above with a triangle. The resulting bound is again in good agreement with the other two methods we used.}. While the metastability constraint involves finding the minimal path connecting the two minima and computing the resulting bounce action numerically, the absolute stability bound is unambiguous. Thus in the left panel of Fig.~\ref{fig:70} we also compare the absolute stability bound, computed analytically, with the one obtained in Ref.~\cite{Hisano:2010re}~\footnote{We extrapolated the results presented in Fig.~4 of  Ref.~\cite{Hisano:2010re}.}, and again find differences similar to those obtained from the metastability bound comparison~\footnote{See footnote \ref{ft:1}.}.

\begin{figure}[t]
\includegraphics[scale=0.85, angle=0]{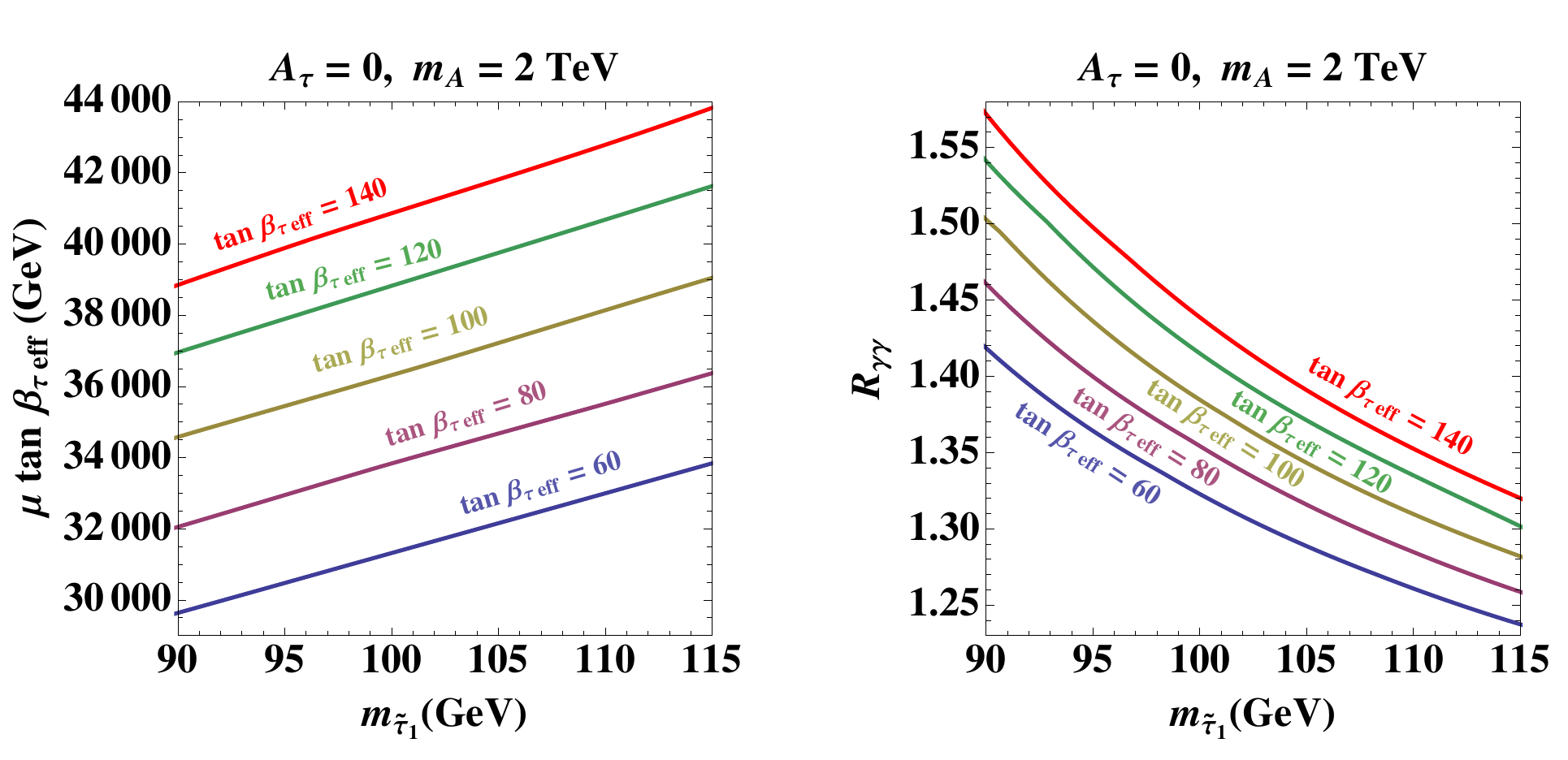} 
\caption{{\em  Left panel: Metastability bound on $\mu\tan\beta_{\tau\,{\rm eff}}$ as a function of $m_{\tilde{\tau}_1}$, for $m_{L_3} = m_{E_3}$, $A_\tau = 0$, $m_A = 2$~TeV, $M_1=55$ GeV, $M_2=400$ GeV and $M_3=1200$ GeV. Right panel: Enhancement in the diphoton partial width  with respect to the SM expectation, as  allowed by the metastability condition,  as a function of the lightest stau mass and for the same supersymmetry parameters as in the left panel.}}
\label{fig:fig3CpsH1}
\end{figure}

Our main goal is to study the $\tan\beta$ dependence of the vacuum metastability bound on $\mu\tan\beta_{\tau\, {\rm eff}}$. {\tt CosmoTransitions} is a public code with the capability of handling the full scalar potential, encompassing both the up- and the down-type Higgses, as well as the one-loop effective potential. Therefore, in the following we choose to present our results based on the outcome of {\tt CosmoTransitions} using the full one loop effective potential, imposing stop mass parameters consistent with a Higgs mass of about 125~GeV. It turns out that using the one-loop effective potential instead of the improved tree-level potential only results in a difference of a few percent on the bound on $\mu\tan\beta$, which explains the small differences between Figs.~\ref{fig:70} and \ref{fig:fig3CpsH1}.

In the left panel of Fig.~\ref{fig:fig3CpsH1}, we present the bound on $\mu \tan\beta_{\tau\,{\rm eff}}$ as a function of the lightest stau mass for different choices of $\tan\beta_{\tau\,{\rm eff}}$. We set $m_{L_3} = m_{E_3}$, $M_1=55$ GeV, $M_2=400$ GeV and $M_3=1200$ GeV. This first plot shows the results in the decoupling limit with no CP-even Higgs mixing from the stau sector: $m_A = 2$~TeV and $A_\tau = 0$. One sees that the bound  becomes weaker as $\tan\beta_{\tau\,{\rm eff}}$ grows. As explained in Section \ref{sect:II},  this is because for a fixed value of $\mu \tan\beta_{\tau\,{\rm eff}}$ the stabilizing quartic term $|\tilde{\tau}_L \tilde{\tau}_R|^2$ increases with  $\tan\beta_{\tau\,{\rm eff}}^2$. In the right panel of Fig.~\ref{fig:fig3CpsH1} we show the allowed enhancement in the diphoton partial width for the same set of parameters. One sees that an enhancement of up to 50\% may be obtained for $\tan\beta_{\tau\,{\rm eff}} \lesssim 100$ and a stau mass of 90 GeV, the LEP limit.  Larger enhancements may be achieved for even larger value of $\tan\beta_{\tau\,{\rm eff}}$~\footnote{The value of $M_1$ we chose is of the right order to generate a proper Dark Matter relic density for values of the lightest stau mass of about 100~GeV~\cite{Carena:2012gp}.  For such values of $M_1$~($2 M_1 < m_h$),  a small invisible width is generated, which becomes more significant for larger values of $\tan\beta$ and $A_\tau$. This enhancement in the invisible width is due to an increase of the lightest neutralino Higgsino component with decreasing values of $\mu$. The total width is then enhanced by at most a few percent in the region of parameters under study, and therefore leads to a reduction  of all visible branching ratios by a similar amount.}.

\begin{figure}[t]
\includegraphics[scale=0.85, angle=0]{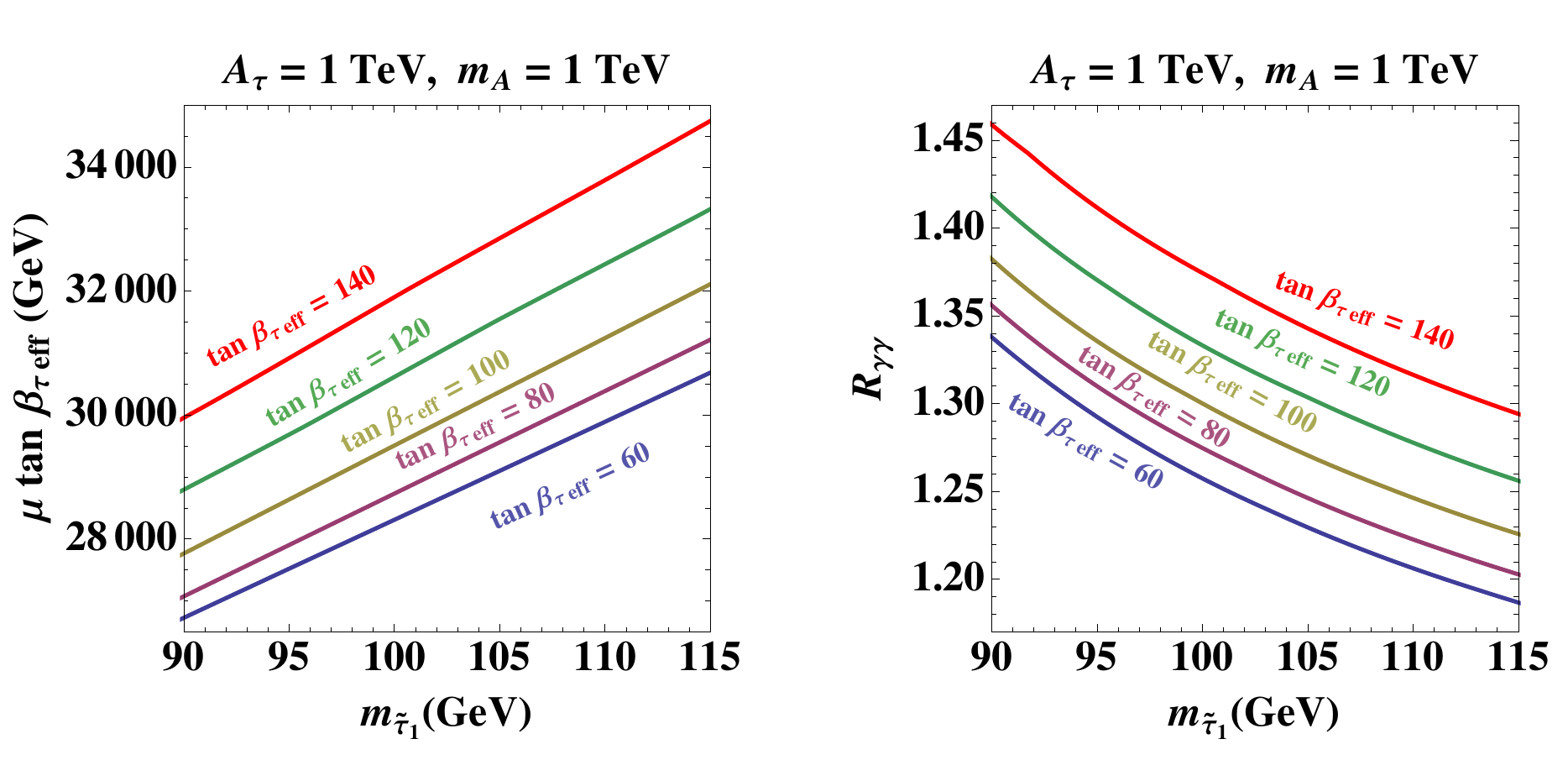} 
\caption{{\em Left panel: Metastability bound on $\mu\tan\beta_{\tau\,{\rm eff}}$ as a function of $m_{\tilde{\tau}_1}$, for $m_{L_3} = m_{E_3}$, $A_\tau = 1$~TeV, $m_A = 1$~TeV, $M_1=55$ GeV, $M_2=400$ GeV and $M_3=1200$ GeV. Right panel: Enhancement in the diphoton partial width  with respect to the SM expectation, as allowed by the metastability condition, as a function of the lightest stau mass and for the same supersymmetry parameters as in the left panel.}}
\label{fig3CPsH2}
\end{figure}

In Fig.~\ref{fig3CPsH2} we show similar plots with $A_{\tau}=m_A=1$ TeV. We again set $m_{L_3} = m_{E_3}$, $M_1=55$ GeV, $M_2=400$ GeV and $M_3=1200$ GeV. Comparing the left panels of Fig.~\ref{fig:fig3CpsH1} and Fig.~\ref{fig3CPsH2}, we note that the bound on $\mu\tan\beta_{\rm eff}$ is about 20\% more stringent  for  the lower  value of $m_A$ and larger value of $A_\tau$.  As discussed before, this is due to the destabilizing effect of the $A_\tau$ trilinear coupling  in Eq.~(\ref{eq:deltaV}), which can generate new charge breaking vacua at  relatively large and negative values of the field $h_d$.

As can be seen from comparing the right panels of Fig.~\ref{fig:fig3CpsH1} and Fig.~\ref{fig3CPsH2}, the effect on the Higgs diphoton decay rate of the more stringent bound on $\mu\tan\beta_{\rm eff}$, in the case of larger $A_\tau$ and smaller $m_A$,  may be partially compensated by the suppression of the $b\bar b$ width as discussed below Eq.~(\ref{ghdd}): an enhancement of the diphoton rate by a factor of $\sim 40\%$ may still be obtained for $\tan\beta_{\tau\,{\rm eff}}\lesssim 100$ and a lightest stau mass at around 90 GeV.
 
\begin{figure}[t]
\includegraphics[scale=0.85, angle=0]{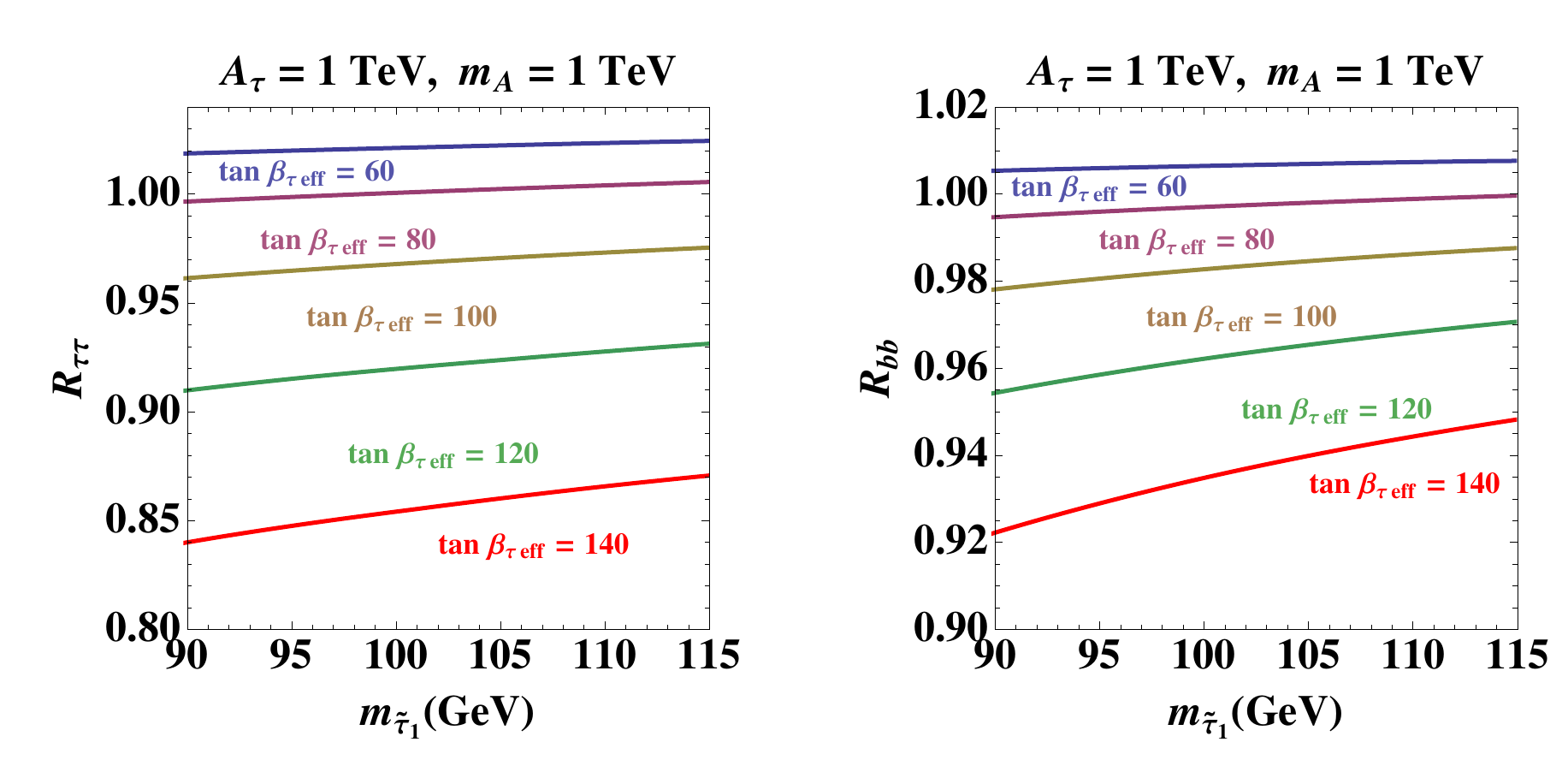} 
\caption{\em Branching ratio of the decay of the lightest CP-even Higgs into{ \em (Left):} two $\tau$ leptons, and {\em (Right):} two $b$ squarks as a function of the lightest stau mass, normalized to  their SM values, for $m_{L_3} = m_{E_3}$, $A_\tau = 1$~TeV, $m_A = 1$~TeV, $M_1=55$ GeV, $M_2=400$ GeV and $M_3=1200$ GeV.}
\label{fig:bbwidthCPsH}
\end{figure}

In order to compute the branching ratio of the Higgs decay into bottom-quarks  and tau-leptons we used the latest version of the public program {\tt CPsuperH}. This program computes the Higgs spectrum and decay rates, as well as  the sparticle spectrum,  including the $\Delta_{\tau,b}$ effects, and therefore provides a consistent framework for computing the Higgs decay widths as a function of the lightest stau mass  within the effective theory approximation described in this work~\footnote{A quantitative comparison of the results obtained from {\tt CPsuperH} with the ones obtained from {\tt FeynHiggs} \cite{Frank:2006yh}  is technically  difficult, since {\tt FeynHiggs} does not include $\Delta_\tau$ effects, but does include $\Delta_b$ effects as well as full one-loop corrections to the $h \tau\tau$ coupling.  Using {\tt FeynHiggs} with $\tan\beta_{\tau{\rm eff}}$ and $A_{\tau{\rm eff}}$  as input yields a similar stau spectrum as the one from {\tt CPsuperH} with the corresponding $\tan\beta$ and $A_{\tau}$, but the bottom Higgs coupling is artificially affected by this change.}. The suppression of the Higgs to $\tau\tau$ and $bb$ rates is shown in Fig.~\ref{fig:bbwidthCPsH}. It is clear from this figure that for these values of $A_\tau$ and $m_A$, a relevant suppression of the $\tau\tau$ rate is only possible for very large values of $\tan\beta_{\tau\,{\rm eff}}$.

Larger values of $A_\tau$ may lead to larger stau contributions to the off-diagonal term of the CP-even Higgs mass matrix element, Eq.~(\ref{loop12}). However,  larger values of $A_\tau$ also induce a stronger metastability bound on $\mu \tan\beta$, which in turn implies that the effect of $A_\tau$ on the Higgs mixing is reduced.  We checked that the combination of all these effects is such that varying the value of $A_\tau$ within a few hundred GeV leads to only small changes on the results presented in Fig.~\ref{fig:bbwidthCPsH}. For increasing values of $A_\tau$, we found that the suppression of $\mu \tan \beta$ was such that  overall there was smaller allowed suppressions of the Higgs to $b\bar{b}$ and Higgs to $\tau \tau$ decay branching ratios. Similar comments apply to variations of $m_A$, although in this case it is smaller values of $m_A$ that may lead to larger CP-even Higgs mixing effects, which are limited by stronger bounds on $\mu\tan\beta$. Therefore, Fig~\ref{fig:bbwidthCPsH} is representative of a more general case and, quite generically, very large values of   $\tan\beta_{\tau\,{\rm eff}}$ are necessary in order for stau effects to modify the Higgs to $b\bar{b}$ and Higgs to $\tau\tau$ decay rates in a relevant way.
   
\section{Constraints on Large {\boldmath $\tan \beta$}}
\label{sec:tb}

\begin{figure}[t]
\subfigure[]{
\includegraphics[scale=0.85, angle=0]{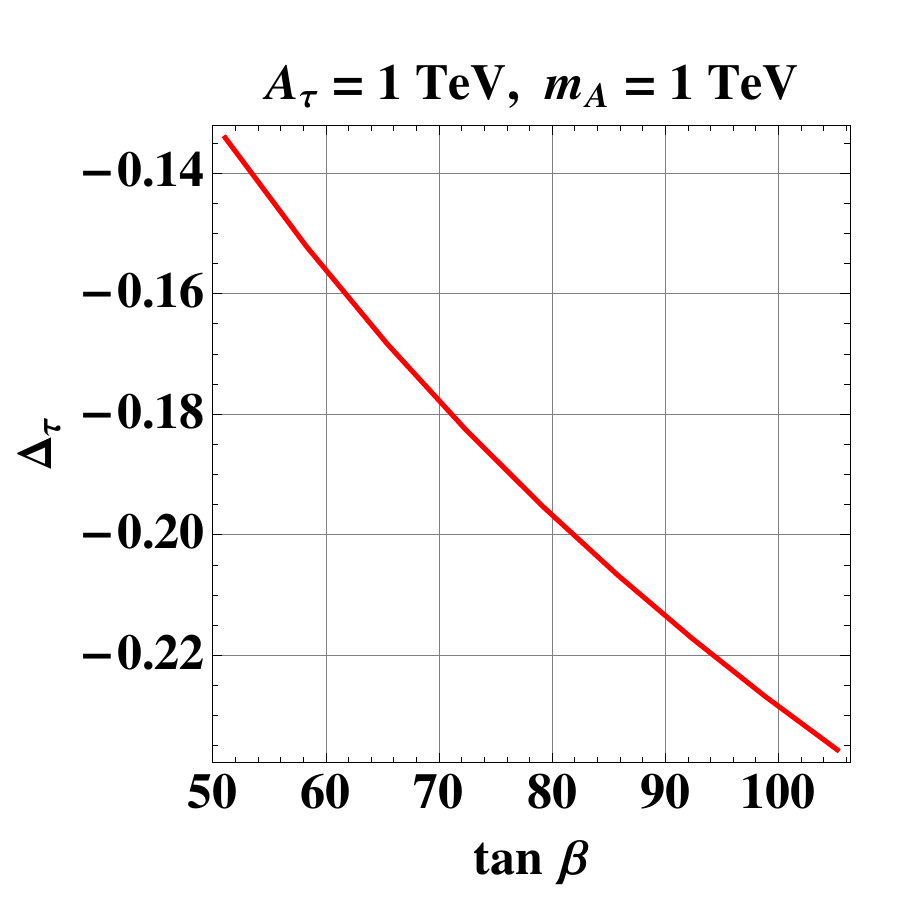} 
}
\subfigure[]{
\includegraphics[scale=0.85, angle=0]{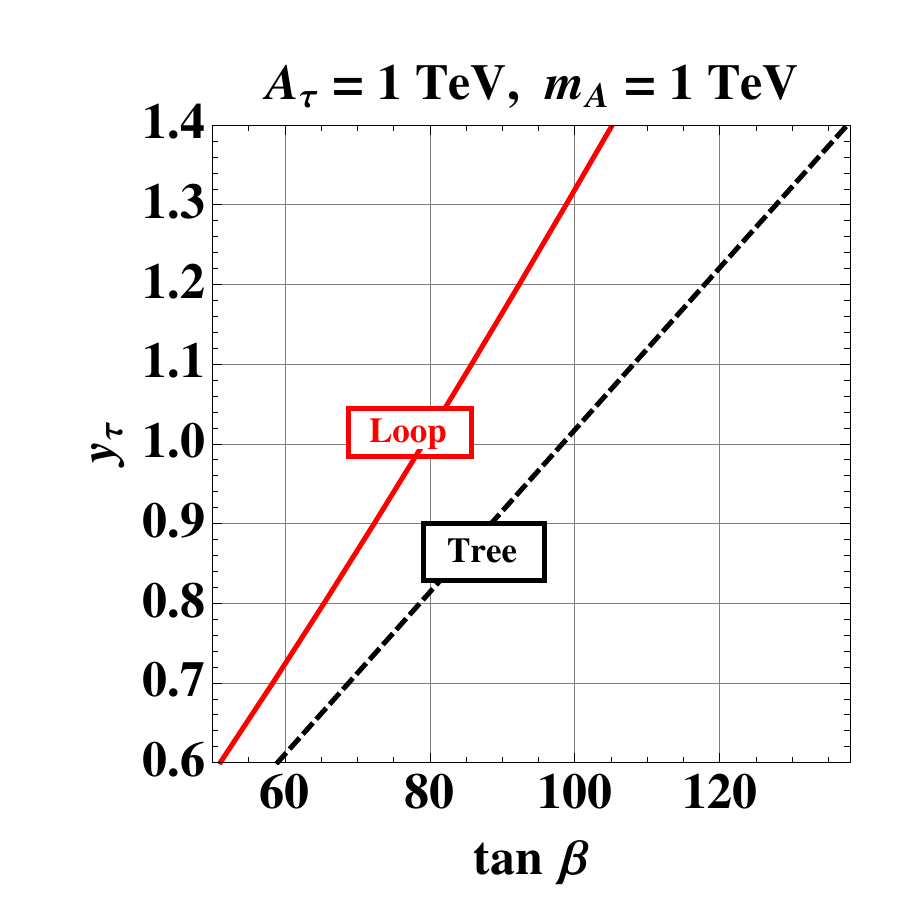}  
}
\caption{\label{fig1}{\em Left panel: $\Delta_\tau$ versus $\tan\beta$ for $m_{L_3}=m_{E_3}=240$ GeV, $M_1=55$ GeV, $M_2=400$ GeV, $M_3=1200$ GeV  and $A_\tau=1$ TeV. The lightest stau mass varies in the range 92-103 GeV, with the lighter masses associated with the larger values of $\tan \beta$. The values of  $\mu$ are taken to be those associated with the metastability limit.  Right panel: $y_\tau$ versus $\tan\beta$ for the same set of parameters. The red~(solid)  line uses the one-loop relation (Eq.~(\ref{eq:looptau})), while the black~(dashed) line uses the tree-level relation (Eq.~(\ref{eq:treeytau})). }}
\label{fig:ytautb}
\end{figure}

\begin{figure}[t]   
\includegraphics[scale=0.9, angle=0]{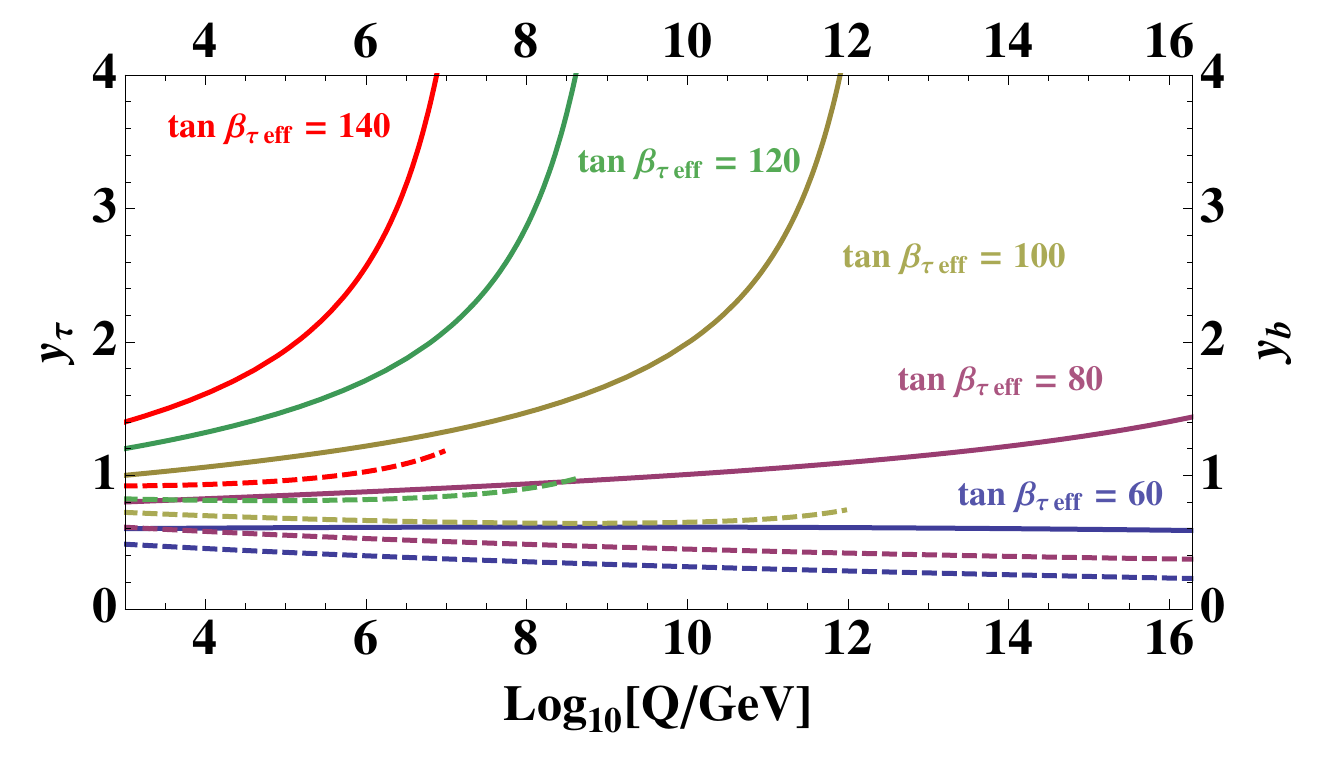}  
\caption{\em Two-loop RG evolution of the $\tau$ (solid lines) and $b$ (dashed lines) Yukawa couplings, for different values of $\tan\beta_{\tau\,{\rm eff}}$.  We terminate the running of the bottom Yukawa at the same scale at which the  tau Yukawa for the corresponding $\tan\beta_{\tau\,{\rm eff}}$ becomes non-perturbative.}
\label{fig:ytaurun}
\end{figure}

The large $\tan\beta_{\tau\,{\rm eff}}$ needed to enhance the diphoton width~(and suppress the $\tau\tau/bb$ couplings) leads to large values of $y_\tau$ and $y_b$ at the weak scale.  One may be concerned that a large tau/bottom Yukawa coupling may become non-perturbative at some intermediate energy scale below the GUT scale, necessitating new physics beyond the MSSM at or below the Landau pole energy. 
 
To get a better sense of the values of $\tan\beta$ and $\tan\beta_{\tau\,{\rm eff}}$ in the region of interest, in Fig.~\ref{fig:ytautb} we plot $\Delta_\tau$ and $y_\tau$ as a function of $\tan\beta$, similar to Fig.~\ref{fig:deltatb}, but this time taking into account the metastability constraints on the $\mu$ parameter. Although in the figure we have chosen specific values  of the  soft supersymmetry breaking slepton masses, we verified that $\Delta_\tau$ is not very sensitive to the resulting stau masses. Therefore, the right panel of Fig.~\ref{fig:ytautb} can be  used to approximately translate between values of $\tan \beta_{\tau\, \rm{ eff}}$ and $\tan \beta$ for the entire range of parameters of  interest in this work. 
 
The  running of the $\tau$ and bottom Yukawa couplings depends strongly on their values at the weak scale, which are determined by $\tan\beta_{\tau \,{\rm eff}}$ and the analogous $\tan\beta_{b \, {\rm eff}}$, respectively. In the region of parameters we are studying, namely soft breaking squark and gluino masses of order 1~TeV and stop mixing $A_t$ of about 1.5~TeV, one obtains $\Delta_b \simeq$~50\%. This leads to values of the bottom Yukawa coupling at the TeV scale of order 0.6 -- 1, rather than the values of order 0.9 -- 1.5 that would have been obtained had we used the tree-level relationship for values of $\tan\beta \simeq$~60 -- 100. In addition, the effect of the QCD coupling on the evolution of the quark Yukawa couplings  decreases their values for increasing energy scales. Thus, rather than from the bottom Yukawa coupling, the strongest perturbativity constraint comes from the running of the $\tau$ Yukawa coupling,  due to its large values at the weak scale induced by a negative $\Delta_\tau$.

Fig.~\ref{fig:ytaurun} shows the two-loop renormalization group (RG) evolution~\cite{Martin:1993zk} of $y_\tau$ and $y_b$ as a function of the RG scale $Q$, using the weak scale Yukawa couplings obtained from including $\Delta_\tau$ and $\Delta_b$ effects. We see that perturbative consistency up to the GUT scale, $\mbox{Log}_{10}[Q/\mbox{GeV}]=16$, may be obtained for $\tan\beta_{\tau \, {\rm eff}} \lesssim 90$, which corresponds approximately to $\tan \beta \lesssim 70$ according to Fig.~\ref{fig:ytautb}. Larger values of $\tan\beta_{\tau \, {\rm eff}}$  demand an ultraviolet completion at scales below the GUT scale due to the appearance of a Landau pole. Comparing these results with the ones presented in Fig.~\ref{fig:bbwidthCPsH}, we observe  that, in the light stau scenario, a suppression of the Higgs decay branching ratio  into $\tau$ leptons larger than $\sim 5\%$ would require an ultraviolet completion of the MSSM at scales below the GUT scale.

Another constraint on the very large $\tan\beta$ regime comes from flavor physics. In the Minimal Flavor Violation hypothesis~\cite{D'Ambrosio:2002ex}, the most important flavor observables receiving $\tan\beta$ enhanced new physics contributions are $B_u \to \tau \nu$, $b \to s \gamma$ and $B_s \to \mu\mu$. These observables could in principle give a stringent bound on the value of $\mu\tan\beta$.  It has  been shown recently that, assuming $A_t>0$ and stop masses of about 1 TeV, the $b\to s\gamma$ branching ratio is enhanced, and the $B_s\to\mu\mu$ tends to be smaller than the SM expectation~\cite{Haisch:2012re}. However, although consistency with the observed  $B_s \to \mu\mu$ value is more easily achieved for $\mu A_\tau >0$, quite generally  the bounds coming from $b \to s \gamma$ and $B_s \to \mu\mu$  depend strongly on the splitting between the third generation and the rest of the squark masses,  an effect not directly related to the Higgs decay rate into diphotons ( see, for example,  Ref.~\cite{Altmannshofer:2012ks}, for a recent discussion). A more robust constraint is instead represented by $B_u \to\tau\nu$: charged Higgs contributions to this decay rate arise at the tree level and are generically large for large values of the tau and bottom Yukawa couplings. It has been shown, however, that, choosing a charged Higgs mass at around 1 TeV would satisfy the bound from $B_u \to\tau\nu$, for the entire range of values for $y_\tau$ and $y_b$ considered in this paper~\cite{Altmannshofer:2010zt}.

Finally, we  note that MSSM heavy Higgs bosons at around 1 TeV are starting to be probed by direct searches through their decays  into $\tau\tau$ and $bb$~\cite{nonstandardHiggs}.  A naive extrapolation of the CMS results presented in Ref.~\cite{nonstandardHiggs} would indicate that values of $m_A \simeq 1$~TeV may be ruled  out a the 95\% C.L. for values of $\tan\beta \gtrsim 70$. However, for such large values of $\tan\beta$ and $A_\tau$, the width of these Higgs bosons is very large. In addition, these heavy Higgs bosons have  relevant decays into stau pairs, which may suppress the branching fraction into the tau leptons \cite{InPreparation}. Therefore, a naive extrapolation may not be valid and a detailed analysis  is necessary in order to determine the constraints on the CP-odd Higgs boson mass from direct searches. 

Other phenomenological constraints, like precision electroweak observables, the anomalous magnetic moment and the dark matter relic density in the light stau scenario were  discussed in Refs.~\cite{Carena:2012gp, Giudice:2012pf} with positive conclusions. These previous studies, however, did not include the effects of $\Delta_\tau$, which are important in this region of parameter space.

 \section{Conclusion}
 \label{sec:Conc}
 
 In this work we studied the vacuum stability constraint on the light stau  scenario in the MSSM, which can produce a significant enhancement in the Higgs-to-diphoton decay width. We improved upon earlier studies of the metastability condition by analyzing in detail the $\tan\beta$ dependence and also including the effects of a non-zero $\Delta_\tau$, which corrects the relationship between the tau lepton mass and the tree-level tau Yukawa coupling.  We find that an enhancement of the order of 50\% is consistent with the requirements of metastability of the electroweak vacuum and  the perturbativity of the MSSM up to the GUT scale. 
 
In the region of parameter space we are interested in, $\Delta_\tau$ is negative and of the order of 10 - 25\%. Hence, the tree-level tau Yukawa, given the measured tau lepton mass, becomes  larger than when neglecting $\Delta_\tau$. Since the stabilizing quartic term of the stau-Higgs scalar potential, $|\tilde{\tau}_L \tilde{\tau}_R|^2$, is proportional to the square of the tau Yukawa coupling, $y_\tau^2$, a larger $y_\tau$ relaxes the bound coming from vacuum stability. Such an effect  becomes more important  for larger values of  $\tan\beta$, and hence it is relevant to take into account the $\tan\beta$ dependence on the metastability constraints on $\mu\tan\beta$. 

In addition, we also studied the impact of having a non-zero $A_\tau$ on both the vacuum stability constraint and the Higgs to diphoton decay width. We found that positive values of $\mu A_\tau$ may significantly impact the metastability condition, lowering the bound on $\mu \tan \beta$ by $\sim 20\%$  for $A_\tau = m_A = 1$~TeV. On the other hand, we also showed that $A_\tau$ impacts the CP-even Higgs mixing and thus can decrease the Higgs decay into taus and bottoms depending on the value of $m_A$.  In particular, for sizable $A_\tau$ and $\mu A_\tau >0$,  the suppression of the Higgs to diphoton width due to lower allowed values of $\mu \tan \beta$ is partially compensated  by the decrease in the total Higgs width, leading to a reduction of no more than $\sim 10\%$ in the Higgs decay rate into diphotons for $A_\tau = m_A = 1$~TeV with respect to  the $A_\tau=0$ case.

We then studied the two-loop RG running of the tau and bottom Yukawa couplings and found that an enhancement in the diphoton width of the order of 50\% is consistent with perturbative values of the Yukawa couplings up to scales of the order of the GUT scale. On the contrary, a significant suppression of the width of the Higgs decay into tau leptons, larger than $\sim 10\%$, requires an ultraviolet completion of the MSSM at scales below the GUT scale.
  
Finally, though more data is necessary to determine if the enhancement suggested by current measurements is real or a product of statistical fluctuations, we have shown that the precise Higgs decay rate to diphotons in the MSSM is intimately connected to the fate of the Universe. In conclusion therefore,  a precise measurement of the Higgs coupling to diphotons has far reaching implications, and should be a high priority at the LHC and at any future Higgs factory.

 \begin {acknowledgements}
 We acknowledge correspondences with Junji Hisano and Carroll Wainwright. We would like to thank Wolfgang Altmannshofer for discussions. Fermilab is operated by Fermi Research Alliance, LLC under Contract No. DE-AC02-07CH11359 with the U.S. Department of Energy. Work at ANL is supported in part by the U.S. Department of Energy under Contract No. DE-AC02-06CH11357. Work at Northwestern is supported in part by the U.S. Department of Energy under Contract No. DE-FG02-91ER40684. Work at KITP is supported by the National Science Foundation under Grant No. NSF PHY11-25915. I.L.  was partially supported by the Simons Foundation under award No. 230683. N.R.S is supported by the DoE grant No. DE-SC0007859.
 \end{acknowledgements}

\section*{APPENDIX}
\label{appendix}
 
 In this appendix we  provide, for convenience, the approximate analytic expressions for $\Delta_\tau$ and $\Delta_b$, the complete expressions given in Ref.~\cite{Carena:2002es}. We first define the loop function 
 \be
 I(a,b,c)= \frac{a^2 b^2 \log(a^2/b^2) + b^2 c^2 \log(b^2/c^2) + c^2 a^2 \log(c^2/a^2)}{(a^2-b^2)(b^2-c^2)(a^2-c^2)} \ .
 \ee
 Then in the region $\tan\beta \gg 1$, one can write
 \bea
 \label{eq:deltatau}
 \Delta_\tau& \simeq&
  - \frac{3 \mu  \tan\beta}{32\pi^2}g_2^2\ M_2\ I(m_{\tilde{\nu}_\tau}, M_2, \mu)- \frac{ \mu  \tan\beta}{16\pi^2}g_1^2\ M_1\ I(m_{\tilde{\tau}_1}, m_{\tilde{\tau}_2}, M_1)
 \ ,  \\
 \label{eq:deltab}
 \Delta_b&\simeq&\frac{\mu \tan \beta}{2 \pi^2} \left[ \frac{g_3^{2}}{3} M_{\tilde{g}} \ I(m_{\tilde{b}_1},m_{\tilde{b}_2}, M_{\tilde{g}}) + \frac{y_t^2}{8} A_t \ I(m_{\tilde{t}_1}, m_{\tilde{t}_2}, \mu)\right] \, .
 \eea
In the above $M_{1(2)}$ is the mass parameter for the Bino (Wino), $m_{\tilde{\tau}_{1,2}}$ are the stau masses,  $M_{\tilde{g}}$ is the gluino mass, $m_{\tilde{b}_{1,2}}$ are the sbottom masses, $A_t$ is the trilinear soft-breaking term in the stop sector, and $m_{\tilde{t}_{1,2}}$ are the stop masses. Moreover, $g_1$ is the gauge coupling for $U(1)_Y$, $g_2$ is the gauge coupling for $SU(2)_L$, $g_3$ is the gauge coupling for $SU(3)_c$, and $y_t$ is the top Yukawa coupling.

\end{document}